# A Sub Pixel Resolution Method

S. Khademi, A. Darudi, and Z. Abbasi

***Abstract***—One of the main limitations for the resolution of optical instruments is the size of the sensor's pixels. In this paper we introduce a new sub pixel resolution algorithm to enhance the resolution of images. This method is based on the analysis of multi-images which are fast recorded during the fine relative motion of image and pixel arrays of CCDs. It is shown that by applying this method for a sample noise free image one will enhance the resolution with $10^{-14}$ order of error.

## I. INTRODUCTION

IN addition to the diffraction and the optical aberrations, the pixel size of an optical detector, $L_{pixel}$, is one of the most important limitations of optical resolution [1]. Often some fine structures of images are smaller than the pixel size and don't recognize by the CCD pixels. To enhance the optical resolution one may use a more expensive detector with more pixels per unit area, or may refer a sub pixel resolution [2-5] or super-resolution methods [6-13], as well as an optical designing engineering for the optical lenses.

In this paper, we arrange an experimental setup to record low resolution multi-images for a two dimensional spatial frequency pattern, during the relative fine motion of the image and the CCD. The images are recorded after each displacement with a length $L_{pixel}/n$, along the row and (or) column of CCD arrays. The analysis of the images in our algorithm will give us an enhanced higher resolution image.

First we show the advantage of our algorithm by applying the method for a sample *low resolution image* which is constructed from a *high resolution reference image*, for the comparison. In this procedure we take the number of steps, $n = 3$ and 5.

The layout of this paper is as follows: In Section 2 we introduce our sub pixel resolution algorithm. In Section 3 the algorithm is applied for a two dimensional noise free image.

S. Khademi is now with the Department of Physics, Zanjan University, 6th Km of Tabriz Road, Zanjan IRAN. Corresponding author to provide phone: 98-241-2538; fax: 98-251-515-2264; e-mail: siamakkhademi@yahoo.com and khademi@znu.ac.ir).

A. Darudi, was with the Department of Physics, Zanjan University, 6th Km of Tabriz Road, Zanjan IRAN. He is now with Lund Observatory, Box 43, SE-221 00 Lund, SWEDEN as a researcher (e-mail: ahmad@astro.lu.se).

Z. Abbasi is graduated in the Department of Physics, Zanjan University, Zanjan IRAN. (e-mail: abbasi.z343@gmail.com).

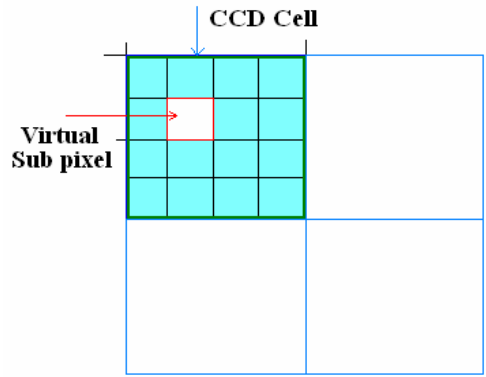

Fig. 1  A CCD pixel (dark green square) is divided into $4 \times 4$ virtual sub pixels (e.g. red square). Blue squares are the zones of image which before any movement is coincided on the individual CCD pixels.

We have a comparison with the reference image and evaluate the error for the enhanced image. We summarize our finding in Section 4.

## II. A NEW PIXEL RESOLUTION ALGORITHM

Supposes a CCD pixel (or cell) virtually divided into $n$ sub-pixels. Figure (1) shows, e.g. 16 virtual pixels (or sub- pixel) in one pixel of CCD. The intensity of each pixel is a total sum of the 16 sub pixels intensity. Thus, the CCD can not recognize the details of the image which are smaller than the pixel size.

The recorded intensity in $(j_x, j_y)$-th pixel and $(m'_x, m'_y)$-th step (see Figs.2-4), is denoted by $I'(j_x, j_y, m'_x, m'_y)$. In this case $m'_{x(y)} = 1,\ldots,n_{x(y)}$, where $n_x$ (or $n_y$) is the number of virtual sub pixels in a pixel in x-direction (or y-direction). After a complete scan of all sub pixels in a pixel and recording the intensities, one obtains N matrices for image $I'(j_x, j_y, m'_x, m'_y)$, where $N = n_x \times n_y$.

Our main aim is to reconstruct the intensity of all sub pixels, which is denoted by $I(m_x, m_y)$, where $m_i$ s are the number of the total sub pixels in the corresponding. Our algorithm is divided in two stages:

*A. Stage 1*

Recording the intensities of CCD pixels, $I'(j_x, j_y, m'_x, m'_y)$, for each step of movements. In this case one scans the first row from $(m_x = 1, m_y = 1)$ to $(m_x = n_x - 1, m_y = 1)$, then the second row from $(m_x = 1, m_y = 2)$ to $(m_x = n_x - 1, m_y = 2)$ and up to the last row from $(m_x = 1, m_y = n_y - 1)$ to $(m_x = n_x - 1, m_y = n_y - 1)$. At the end of stage 1, one has $N$ images[1].

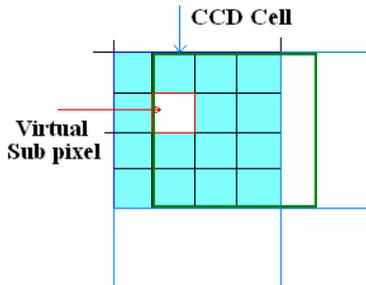

Fig. 1 First step of movement of CCD pixels in x-direction. In this figure the displacement (or sub pixel size) is equal to one forth of the pixel size.

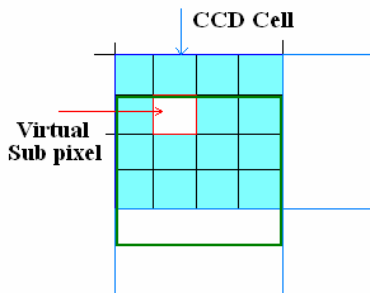

Fig. 3 First step of movement of CCD pixels in y-direction.

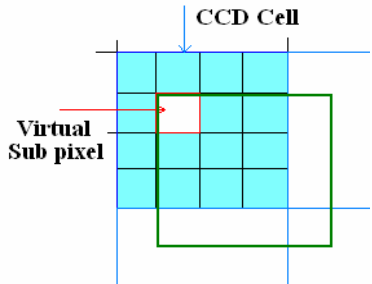

Fig. 4 The CCD pixels are moved one step in y-direction and one step in x-direction.

Figures (5) and (6) show the position of image and CCD pixels for no movement $(m_x = 1, m_y = 1)$ and x-directional movement $(m_x = 2, m_y = 1)$, respectively. Figure (5) shows the margin condition for our algorithm where a CCD pixel column, $j_x = 0$, is out of image zone (in darkness) and has zero intensity. This marginal condition for sub pixel resolution method has an important role to reconstruction the images. There is a similar marginal condition for $j_y = 0$, a row which initially is out of image in darkness. The first stage is divided into four parts:

---

[1] Note that $(m_x = n_x, m_y = n_y)$ is the same as $(m_x = 1, m_y = 1)$ for another pixel, therefore it is a repetitive record image which is neglected.

*Part-1:* For the up left corner (first) pixel of CCD (where $j_x = 1, j_y = 1$) one finds the intensity of pixel as

$$I'(j_x, j_y, m'_x, m'_y) = \sum_{i=1}^{m'_x} \sum_{k=1}^{m'_y} I(i,k). \quad (1)$$

*Part-2:* The intensity for the next pixels in the first row, $j_x = 1, 2, \cdots, j_y = 1$, is given by

$$I'(j_x, j_y = 1, m'_x, m'_y) = \sum_{i=(j_x-2)n_x+m'_x+1}^{(j_x-1)n_x+m'_x} \sum_{k=1}^{m'_y} I(i,k). \quad (2)$$

*Part-3:* The intensity for the next pixels in the first column, $j_x = 1, j_y = 1, 2, \cdots$, is given by

$$I'(j_x = 1, j_y, m'_x, m'_y) = \sum_{i=1}^{m'_x} \sum_{k=(j_y-2)n_y+m'_y+1}^{(j_y-1)n_y+m'_y} I(i,k). \quad (3)$$

*Part-4:* Equations (2) and (3) give the intensity of pixels in the first row and column of CCD pixels. For the intensity of other pixels of CCD one has

$$I'(j_x, j_y, m'_x, m'_y) = \sum_{i=(j_x-2)n_x+m'_x+1}^{(j_x-1)n_x+m'_x} \sum_{k=(j_y-2)n_y+m'_y+1}^{(j_y-1)n_y+m'_y} I(i,k). \quad (4)$$

Three parts of the first stage define the intensities which is recorded in the first column and row of image. It gives us the initial conditions which is necessary for the second stage of the algorithm. The non-zero intensity for the out of image pixels will generate some error in our algorithm.

*B. Stage 2*

After the recording of the intensity of pixels in all steps of movement of CCD, we are in a position to reconstruct the sub pixels intensity in the second stage. To reconstruction of the enhanced image follow the following procedure. This reconstruction method is divided into 10 parts:

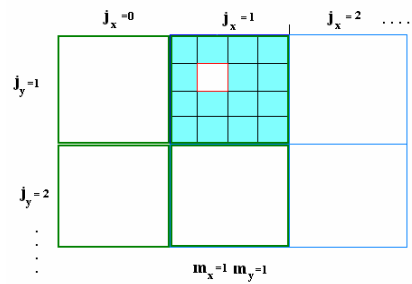

Fig. 2 The intensity $I'(1,1,0,0)$ which is recorded for the pixel in 1st row and 1st column where there is no step movement.

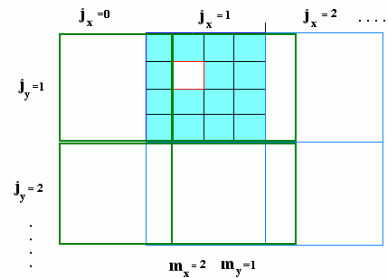

Fig. 3 The first movement of CCD to the x-direction. The first column of CCD at first is in darkness (out of image zone) and is necessary for the marginal condition in the next steps of movements.

*Part-1:* The number of pixels in a row (and column) is $N_x$ (and $N_y$) and the number of sub pixels in a row (and column) is given by $m_x = N_x \times n_x$ (and $m_y = N_y \times n_y$). It is necessary to find the number of row and column of the pixel which contains the $(m_x, m_y)$-th sub pixel. The $(m_x, m_y)$-th sub pixel is in a pixel which is denoted by $(j_x, j_y)$, where

$$j_x = [\frac{m_x}{n_x}] + 1, \quad (5)$$

$$j_y = [\frac{m_y}{n_y}] + 1, \quad (6)$$

and $\frac{m_x}{n_x} = [\frac{m_x}{n_x}] + \frac{m'_x}{n_x}$. In this relations $[p]$ is denoted to the integer part of $p$. Therefore one has

$$\begin{cases} m'_x = m_x - n_x(j_x - 1) \\ m'_y = m_y - n_y(j_y - 1) \end{cases} \quad (7)$$

for $m'_x, m'_y \neq 0$.

*Part 2:* Due to the initial condition the intensity of first sub pixel is given easily by:

$$I(m_x = 1, m_y = 1) = I'(j_x = 1, j_y = 1, m'_x = 1, m'_y = 1). \quad (8)$$

*Part-3:* For the first sub pixels on the first column $m_x = 1$ and $1 < m_y < n_y$ one has

$$I(m_x = 1, m_y) = I'(j_x = j_y = 1, m'_x = 1, m'_y = m_y) - \sum_{k=1}^{m_y - 1} I(1, k). \quad (9)$$

*Part-4:* For the other sub pixels on the first column $m_x = 1$ and $m_y \geq n_y$ one has

$$I(m_x = 1, m_y) = I'(j_x = 1, j_y, m'_x = 1, m'_y) - \sum_{k=m_y-n_y+1}^{m_y-1} I(1, k). \quad (10)$$

*Part-5:* For the first sub pixels on the first row $m_y = 1$ and $1 < m_x < n_x$ one has

$$I(m_x, m_y = 1) = I'(j_x = j_y = 0, m'_x = m_x, m'_y = 1) - \sum_{i=1}^{m_x - 1} I(i, 1). \quad (11)$$

*Part-6:* For the other sub pixels on the first row $m_y = 1$ and $m_x \geq n_x$ one has

$$I(m_x, m_y = 1) = I'(j_x, j_y = 1, m'_x, m'_y = 1) - \sum_{i=m_x-n_x+1}^{m_x-1} I(i, 1). \quad (12)$$

*Part-7:* For the other sub pixels in the first pixel $1 < m_x < n_x$ and $1 < m_y < n_y$ one has

$$I(m_x, m_y) = I'(j_x = j_y = 1, m'_x = m_x, m'_y = m_y) \\ - \sum_{i=1}^{m_x-1} \sum_{k=1}^{m_y} I(i, 1) - \sum_{k=1}^{m_y-1} I(m_x, k). \quad (13)$$

*Part-8:* For the other sub pixels which $m_x \geq n_x$ and $1 < m_y < n_y$ one has

$$I(m_x, m_y) = I'(j_x, j_y = 1, m'_x, m'_y) - \sum_{i=m_x-n_x+1}^{m_x-1} \sum_{k=1}^{m_y} I(i, 1) \\ - \sum_{k=1}^{m_y-1} I(m_x, k). \quad (14)$$

*Part-9:* For the other sub pixels which $m_y \geq n_y$ and $1 < m_x < n_x$ one has

$$I(m_x, m_y) = I'(j_x = 1, j_y, m'_x, m'_y) - \sum_{i=1}^{m_x-1} \sum_{k=m_y-n_y+1}^{m_y} I(i, k) - \sum_{k=m_y-n_y+1}^{m_y-1} I(m_x, k). \quad (15)$$

*Part-10:* For the other sub pixels which $m_y \geq n_y$ and $m_x \geq n_x$ one has

$$I(m_x, m_y) = I'(j_x, j_y, m'_x, m'_y) - \sum_{i=m_x-n_x+1}^{m_x-1} \sum_{k=m_y-n_y+1}^{m_y} I(i, k) \\ - \sum_{k=m_y-n_y+1}^{m_y-1} I(m_x, k). \quad (16)$$

This procedure gives us the sub pixel resolution of our image $I(m_x, m_y)$.

### III. RESULTS FOR A SAMPLE IMAGE

To apply our sub pixel resolution method for a 2D image, one use a noise free sample image in Fig. 7 as a reference image. Its corresponding low resolution image is shown in Fig. 8 and application of reconstruction procedure, by a MATLAB programming, gives the enhanced image in Fig. 9. It shows the reconstructed image is coincides to the reference image by a high accuracy. Figure 10 shows the difference of an arbitrary row between the intensity of the reference and enhanced image, which is in order of $10^{-14}$.

In this procedure the number of sub pixels in a square pixel is assumed to be 100. Our algorithm has some sensitivity to the noises of optical detectors. Therefore this method should be applied for the noise free images or one may apply it simultaneously by other denoising algorithms.

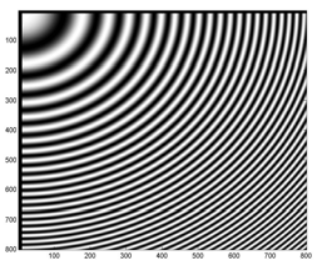

Fig. 7 The two dimensional sample reference image.

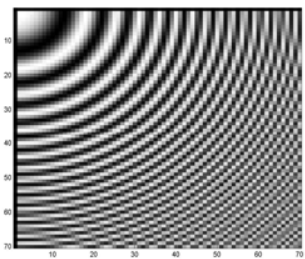

Fig. 8 A simulation of the recorded intensity of Fig. 7 for a low resolution CCD.

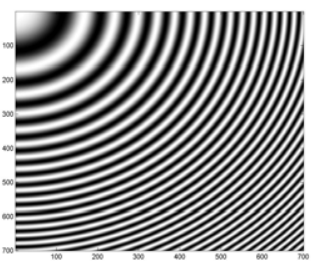

Figure 9: The enhanced image which is obtained by applying our algorithm on the low resolution image Fig. 8.

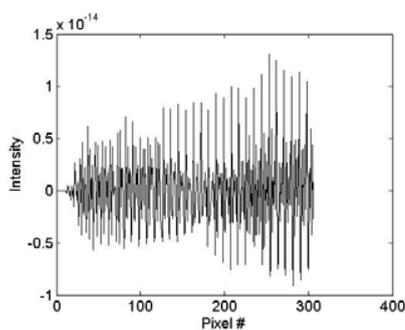

Fig. 10 The difference between the reference and reconstructed images. The magnitude of their difference is about $10^{-14}$ which show the ability of our sub pixel resolution method.

## IV. CONCLUSIONS

In this paper, the algorithm of a sub pixel resolution method for two dimensional images is presented. We show the high coincidence of reference and reconstructed images. The accuracy of our algorithm is in order of $10^{-14}$ which is negligible in many applications. This algorithm is noise dependent and is not an efficient method for noisy images. By our sub pixel resolution method one can take a high resolution picture by a low resolution CCD. Developing our methods to noise friendly algorithms is open now.